# On the Use of Vickrey Auctions for Spectrum Allocation in Developing Countries

**G. Anandalingam**[*]
Robert H. Smith School of Business
University of Maryland
College Park, MD 20742
ganand@rhsmith.umd.edu

and

Systems Engineering and The Wharton School
University of Pennsylvania
Philadelphia, PA 19104

September 2001

Abstract: In this paper, we assess the applicability of auctions based on the Vickrey second price model for allocating wireless spectrum in developing countries. We first provide an overview of auction models for allocating resources. We then examine the experience of auctioning spectrum in different countries. Based on this examination, we posit some axioms that seem to have to be satisfied when allocating spectrum in most developing countries. In light of these axioms, we provide a critical evaluation of using Vickrey second-price auctions to allocate spectrum in developing countries. We suggest the use of a new auction mechanism, the Vickrey "share auction" which will satisfy many of these axioms.

---

1.  Introduction

Many countries are in the process of privatizing their telecommunications systems, and are trying to leapfrog technology by speeding up the investment in wireless communications. Bandwidth and right-of-way are some of the most important public property rights in expanding telecommunications infrastructure. In most countries, right-of-way whether for wireless or wireline systems is controlled by either local or central governments. In the U.S. for instance, there are stringent local municipal rules about when and where new telecommunications equipment and networks can be installed physically. All countries control the airwaves and only allow companies to use it under strict rules. Many of these countries are in the process of allocating the wireless spectrum to private operators of telecommunications networks. In this paper, we examine the issues involved with allocating spectrum, critique the current methods used in countries like India, and use the experience of a number of countries and economic theory to propose a new model for spectrum allocation that could work.

**1.1. Auctions as a Method for Allocating Spectrum**

R.H. Coase in his article, *The Federal Communications Commission*, makes a persuasive argument why the most efficient spectrum policy is to define spectrum as property and to let the market decide what it is worth. In comparing frequency policy to private property, Coase states:

> "Land, labor, and capital are all scarce, but this, of itself, does not call for government regulation. It is true that some mechanism has to be employed to decide who, out of the many claimants, should be allowed to use the scarce resource. But the way this is usually done in the American economic system is to employ the price mechanism, and this allocates resources to users without the need for government regulation" (See Fritts, 1999)

One of the best available methods for pricing the property rights is to allow the market to decide it. However, because spectrum is not a commonly traded commodity, an auction mechanism is the best way of discovering the price for the property-right. A successful auction of spectrum should result in an efficient distribution of this scarce resource to those who are willing to pay the highest price. This is announced as the aim of an auction as defined in the FCC report to Congress on Spectrum Auctions 1997, quoted in Brian C Fritts (1999). A properly designed auction would give incentive for bidders to reveal their true valuation for the spectrum licenses. This would result in allocating licenses to those that value the licenses the most, and thus place the licenses in the hands of firms that would be able to most efficiently use them.

Moreover, the government can also use the auction to promote objectives like development and deployment of technology and services in a speedy fashion, and use spectrum licenses to spur economic development and competition, and recovery of the commercial value of the spectrum. This is listed as the US Congress objectives for the PCS auctions (Fritts, 1999). Setting up an auction would lead to better and more cost effective use of spectrum for all sorts of current and emerging telecommunications technology that in the end would provide consumers with cheaper prices and better service. With the use of auctions the government would be expected to generate more revenue than with other methods; there was also the argument that revenue generated from auctions would be much less distortionary than revenue collected via taxes because the competitors would bid their true economic value.

### 1.2. Experience in Developed Countries in Auctioning Spectrum

The movement towards deregulation of telecommunications industry worldwide has recently manifested in the assignment of airwave spectrum via auction mechanisms. Countries such as the U.S., England, Australia, New Zealand, and Canada have held auctions to assign airwave licenses to bidders. In addition, developing countries such as India are just beginning to get involved in auctioning off spectrum in order to build out their wireless communications infrastructure. The spectrum was initially allocated using hearings and lotteries. Next, most of these countries started auctioning off the spectrum the same way in which many public infrastructure goods were put out for "tender". The interested parties had to submit bids in response to RFPs (Request For Proposals) wherein they delineated their qualifications of being able to build networks, gave plans on how and how soon they would build the network, and then provide the monetary value of the actual bid for the license including some payment schedule. This sealed-bid first-price auction was used to choose the entity to which the license should be allocated. Finally, most developed countries started using a Vickrey auction mechanism, and a variation of it called the ascending price auction mechanism, similar to the English auction used to sell precious art.

New Zealand's experience of using Vickrey auctions provides the starkest evidence of one of its major problems. Starting from 1989, the New Zealand spectrum auctions were of the Vickrey second-price sealed bid variety. In the 1990 auction, the winners of some licenses were essentially getting them for free due to the fact that the second highest bid would be very low. In one of the 1990 auctions in New Zealand, the winning firm's bid was $100,000, but the second-price was $6 (in New Zealand dollars)! Starting from 1991 and continuing to 1994 spectrum auctions in New Zealand went back to the first-price sealed bid format. Thus, it appears that the Ministry of Economic Development in New Zealand decided that it would rather have winners possibly exposed to the winner's curse rather than having reduced revenues. Starting from 1995 the New Zealand Ministry adopted an ascending simultaneous multiple round auction much like the one used by the FCC in the U.S. spectrum auctions

The Federal Communications Commission (FCC) of the United States pioneered the use of the simultaneous ascending-price multiple round (SAMR) auction format wherein the bidders were allowed to simultaneously bid for different resources (eg. contiguous spectrum) and keep raising their bids in subsequent rounds at increments set by the auctioneer until there were no more bids at the last price. The SAMR is a form of Vickrey pricing because the winner only pays an increment above the second-price. (We will elaborate further on all auction mechanisms in section 2). The countries that used this format had also adopted the additional constructs of activity rules, minimum bid increments, designated categories, and stopping rules etc. to enable the auction to work effectively and efficiently. This is not to say that the auctions were held problem free, evidence of collusion was present and it is of course impossible for communications ministries to get true efficiency results since firm valuations are not available.

The traditional method of assigning airwave licenses in the U.S. had been through lotteries, or comparative assessment hearings. (Details of the U.S. history in allocating spectrum can be found in FCC (2001)). Through comparative hearings the FCC would employ an administrative judge to decide which prospective license holder was most deserving of a license. This process was afflicted by inefficiencies in assigning licenses. It would take too long to decide the license assignments among the very many that applied and as a result there would be a big backlog of unassigned licenses. In addition, firms would often have to spend even more time after the hearings trying to aggregate particular licenses from different license holders in order to form geographical synergies in their telecommunications networks. As a result, there was considerable delay of new communications services offered to customers due to these extra dealings.

The FCC next used a lottery mechanism to assign winners of licenses in a more timely fashion. However, this attracted many applicants for licenses, as many small telecommunication firms were attracted by the possibility of winning a valuable spectrum license it otherwise would have no chance of obtaining. It was not uncommon for small companies to win licenses and sell them to larger telecommunications firms afterwards for much larger amounts than they paid the government. Again, there were inefficiencies due to time delays. Sometimes, the smaller companies could not resell their licenses, and had to go into bankruptcy because they could not afford to pay the license fees (for example, NextWave Communications). Thus, from 1994 the FCC decided to abandon these methods in favor of an auction mechanism.

Over five days in July 1994 the FCC auctioned off ten nationwide narrowband and PCS licenses. The primary use of spectrum associated with these licenses was for advanced paging and data services. This auction offered the largest amount of spectrum for sale and lasted 47 rounds over *five* days. There were initially 29 bidders who submitted upfront payments and in the end six of the bidders were to claim the ten licenses. The government collected $617 million for the ten narrowband licenses through the auction.

Next the FCC auctioned regional narrowband PCS licenses in October and November of 1994. The FCC designated five regions for which six licenses would be available per region. Unlike, the nationwide narrowband auction, geographical aggregation would be a critical issue in this auction as firms would be competing for multiple licenses to form synergies. The auction started with 28 bidders for 30 regional licenses and nine firms won the licenses. It took 105 rounds and the government collected $395 million dollars in revenue. The MTA broadband auctions began on December 5, 1994. This auction would be the largest public auction and most important spectrum

auction ever with 99 licenses being auctioned, and 30 bidders (including some of the largest telecommunications firms in the world) competing for licenses whose sale culminated in over $7 billion in revenue. The FCC continues to use SAMR auctions to this day.

Australian government had some adverse experience in the auctioning of licenses for satellite-television services. A first-price sealed bid auction was used to auction two satellite-television services in April 1993. The winners of the two licenses defaulted on their bids, and caused a cascade of defaults until the next highest bidder was found that could pay for his/her bid. The final prices actually paid were on the order of 100 million dollars less than the original bids. The Australian government then decided to use a simultaneous ascending-price multiple-round (SAMR) auction mechanism (with bid default penalty mechanisms) in 1997, and continues to use this auction format predominantly for spectrum license assignment.

In 1997, 838 spectrum licenses, ranging from 12.5 Khz to 1 Mhz, were put up for auction using the simultaneous ascending format. In April and May 1998 the PCS auction raised 350 million dollars in revenue with 2.7 million dollars in withdrawal penalties. The second PCS auction in September 1998 raised 30.631 million dollars. The auction held in February 1999 ended in 38 rounds and raised 66.2 million dollars (one bidder, AAPT LMDS Pty Ltd, claimed all licenses). The third PCS auction was similar to the second: only one bidder registered and this sole bidder paid the reservation price of $20,000 to obtain the license. Motorola Australia Ltd bid $47,000 to win both licenses in the April 1999 auction. Finally, the April 2000 1.8 Ghz PCS auction ended in 138 rounds and raised 1.32 billion dollars.

The United Kingdom held its first auction for spectrum in March 2000 by using a simultaneous ascending auction for five spectrum licenses for $3^{rd}$ generation wireless communications. These licenses would be some of the most sought after licenses because access to the associated spectrum would enable the creation of many innovative mobile communications devices that will soon dominate the telecommunications industry e.g. mobile internet. The largest license would be reserved for a new entrant into the UK telecommunications market.

The decision to use a simultaneous ascending auction was based on the successful experience of the FCC in their use of the auction format. Thirteen bidders competed for the 5 licenses and five bidders won five licenses in this UK spectrum auction that lasted 150 rounds spanning two months. The reserve price for the five licenses was five hundred million pounds sterling. Collectively, the five licenses netted over 22 billion pounds.

Canada also has been a recent entrant into the world of spectrum auctioning. Like most countries, Canada used to assign spectrum licenses through administrative hearings but soon came to realize the inefficiencies of such an approach. Like other countries Canada has also adopted the simultaneous ascending auction format pioneered by the FCC. Recent auctions involved the auctioning of licenses in the 24 to 28 Ghz band. 12 bidders competed for 260 licenses and the auction revenues totaled 171 million dollars.

The auction *lasted 24 days* (Oct. 18, 1999 to Nov. 19, 1999) and went through 117 rounds.

What is clear from these cases is that the simultaneous ascending-price multiple round auctions are becoming the most popular way to allocate spectrum in most developed countries. Developing countries work under conditions that are different enough for one to take a closer look at Vickrey auctions and Vickrey based SAMR auctions to see if they are really applicable. This is the main objective of this paper. While the auctioneer in a developed country designs the auction in order to maximize revenue, at a minimum the governments in most developing countries who are far behind in the telecommunications revolution would like to have networks in place, whether or not they get the maximum license revenues. Further, any auction that takes too long or involves too complicated a set of calculations will allow non-transparent corrupt practices to enter into the process. We will look at the particular conditions in a developing country, and use the case of India's foray into telecom privatization to propose axioms that have to be satisfied by any auction mechanism that would work in these places. We will then present a novel idea based on Vickrey auctions that may have potential.

### 1.3. Organization of the Paper

The paper is organized as follows: In the next section, we describe the three most popular auction mechanisms and discuss the pros and cons of each. In section 3, we give details on India's case in auctioning off telecom property. In section 4, we propose some axioms for conducting auctions in developing countries. In section 5, we discuss the limitations of the Vickrey and SAMR auction in satisfying these axioms. In section 6, we present the Vickrey share auction. We end the paper with concluding remarks in section 7.

## 2. Auction formats

We will first describe different auction formats that have been used for spectrum allocation, and discuss the pros and cons of each (See also Agorics, 1996). In conjunction with the international practice that we described earlier, we will see why Vickrey auctions, especially the ascending price Vickrey auctions, have good theoretical properties that recommend them. In the next section, we will describe the experience of using these different auction formats in a developing country like India.

### 2.1. First-Price Sealed-Bid Auctions:

The simplest method and probably the most popular is the sealed bid format where different telecommunications companies would bid for the right to use the airwaves using a single sealed bid. This method is widely practiced in a number of industries, especially in civil engineering works, and was the methodology of choice used in the first set of cellular auctions in the telecommunications sector in India.

There are two main advantages of the first-price sealed-bid auction: First, it is simple, and merely involves all interested parties submitting a single bid for the resource that is being auctioned. Secondly, it is economically efficient in that the resource, in this case the license to use the spectrum, is given to the one who claims to value it the most.

There are a number of problems also. There is no guarantee that the bidders will reveal their true value for the resource. Thus, it may well be that a bidder who does not truly value the resource the highest will obtain it. There is significant pressure to do "bid shading", i.e. to lower the bid below the actual value of the resource to you. The entity winning the bid might be able to sell to a losing entity which values the resource higher but "shaded" their bid too much. Thus, the auctioneer may obtain a lower revenue than if the bidders bid truthfully. Although in most auctions the rules of the game make collusion illegal, there may be incentives to collude and lower the bids, and subsequent to the auction for the bidders to engage in a secondary market.

Another major problem of a sealed bid "auction" is what economists' call the "Winner's Curse". In a first-price sealed bid auction, it is often the case that the winner may far more than it should; e.g. the difference between the winning bid and the second highest bid would often be quite large. With this scenario, the auctioneer gets very high revenue if the winner follows the rules of the game. However, the winner of the auction may either balk at paying too much, or may be unable to built out the network having paid too much for the license. In some cases, the winner might actually go bankrupt (for example, NextWave Communications in the U.S. that we mentioned earlier) by borrowing a lot to bid high, but by not being able to service this debt in the loan period through revenue generation.

Note that the first-price sealed-bid auction is similar to what is called a "Dutch auction". In a Dutch auction, used to sell tulips, the auctioneer starts with a given (usually high) price for the resource and then progressively lowers its asking price. The person who cries out the first bid for a resource gets it; the one with the highest or first price gets the resource being auctioned.

**2.2. Sealed-Bid Vickrey Auctions:**

In order, to design truth-revealing auctions, Vickrey (1961) came up with a second-price auction. Sealed-bid Vickrey auctions are those where the winner of an auction gets the resource but only has to pay the price of the second highest bidder to obtain it. Second-price auctions like the Vickrey auction are supposed to reduce this winner's curse by allowing the winner of the auction to pay the second highest bid. Like the first-price auction, the bids are sealed, and each bidder is ignorant of other bids unless there is collusion. The item is awarded to highest bidder at a price equal to the second-highest bid (or highest unsuccessful bid). In other words, a winner pays less than the highest bid. If, for example, bidder A bids $10 million B bids $15 million and C offers $20 million, bidder C would win, however he would only pay the price of the second-highest bid, namely $15 million.

Vickrey auctions are both economically efficient and truth revealing. It is easy to show that, because the winning bidder needs only pay the second highest bid, there is no incentive to "cheat" and misrepresent the true value of the resource. Additionally, in a Vickrey auction, the highest bidder always gets the resource, or spectrum in this case; thus, it is economically efficient.

One wonders why any seller would choose this method to auction goods. It seems obvious that a seller would make more money by using a first-price auction, but, in fact, that has been shown to be untrue (Agorics, 1996). Bidders fully understand the rules and modify their bids as circumstances dictate. In the case of a Vickrey auction, bidders adjust upward. No one is deterred out of fear that he will pay too high a price. Aggressive bidders receive sure and certain awards but pay a price closer to market consensus. The price that the winning bidder pays should theoretically be determined by competitors' bids alone and does not depend upon any action the bidder undertakes. Less bid shading or collusion would occur because people don't fear Winner's Curse, and thus the seller might well receive higher revenue in a Vickrey auction. Myerson (1981) proved that the *expected* revenue under both a sealed-bid first-price auction and a Vickrey auction (sealed-bid or English) was the same; i.e. the *revenue equivalence* theorem.

Of course, there is nothing to prevent a general sense of risk aversion in which all bidders would "shade" their bids. In this case, certainly the auctioneer will receive lower revenue. The auctioneer can be insincere and try to increase the value of the second highest bid. This can be done in two ways: Either overstate the second highest bid without actually revealing it, or else get a proxy to bid high and only allocate the resource if there is a bid higher than the proxy bid. Neither is usually allowed in legal Vickrey auctions, and a number of mechanisms are used to ensure that the auctioneer does not behave in an insincere manner. For a sealed-bid Vickrey auction to be above suspicion, the auctioneer, at the end of the auction, will have to reveal the value of all the bids and also the identity of all who made them.

**2.3. Simultaneous Ascending Multiple Round Auctions**

Another auction mechanism that takes care of the issue of transparency, and also fits the Vickrey second-price framework is the simultaneous ascending multiple round (SAMR) auction. The bidders bid on the resource in a round, and the auctioneer then raises the minimum bid by an increment and allows the bidders to bid again in the subsequent round. The auction stops when no bidder accepts the auctioneer's increment and revises his bid upward. This is very similar to the English auction. The main difference between a SAMR auction and an English auction is that the latter is an open-cry auction while the former need not be. SAMR auctions are economically efficient in that the highest bidder gets the resource, and is Vickrey in the sense that the winning bidder obtains the resource at the second-best price plus a small increment. Thus, the SAMR auction is incentive compatible in that, theoretically, there is no incentive to bid untruthfully. One further nicety of the auction is that the winning bidder neither has to reveal the exact value of the resource to it all at once, nor even at the end because the auction stops after an incremental bid above the second highest, bid.

The ascending nature of the simultaneous ascending auction is supposed to help bidders dynamically bid for various licenses as a bidder can re-evaluate future bids in up and coming rounds by utilizing information from the bids placed by other bidders during a current round. Thus, the winner's curse can be reduced by the observation of competing bids during a round i.e. it would be unlikely for a bidder to bid in tremendous excess of a current high bid by accident. In addition, the ascending aspect does not allow bidders to submit new bids that are lower than previous bids. This feature allows for more revenue generation, which is obviously beneficial from the government's point of view.

As described in the Introduction, the FCC of the United States became the first regulator to use SAMR auctions to allocate spectrum at the recommendation of leading economic theorists. SAMR auctions are also much more easily amenable to the auctioning off of multiple licenses, thus allowing bidders to bid on combinations of licenses that will form synergies e.g. licenses that correspond to contiguous geographic regions will allow firms to have economies-of-scale in serving a larger region not to mention a larger customer base. Also, synergies may be formed with existing licenses that firms own, and so synergies may come from sets of licenses up for auction that do not necessarily form contiguous geographic coverage by themselves, but do so in combination with existing licenses owned by firms. So the logic here is that firms would obviously take these factors into consideration when deciding how to value and bid for different licenses and thus a high valuation for a set of licenses would reflect that these licenses would add great value to the firm. Thus, given that highest bidders win auctions then it would be very beneficial to use auctions to allocate these scarce resources (spectrum) so that firms that are highest bidders (i.e. firms that would most likely make most efficient use of spectrum) would end up with the licenses.

SAMR auctions do not necessarily prevent collusions whereby all of the bidders bid low. In order to ensure some level of revenue, the auctioneer might actually post a reservation price and only accept bids above it. This practice is quite common in the English auctions used to sell precious art in Sotheby's and Christie's.

### 3. Telecom Auctions in Developing Countries: The Case of India

In India, the opening up of the telecommunications services sector to full competition began in earnest in the early 1990's. The paradox of millions of unfulfilled applications, abundance of inexpensive telecommunications equipment and no new telecom services prodded the government as well as the political intelligentsia to rethink the fundamentals of the sector (EPW, 2000). Moreover, there was a growing realization among policy makers that good telecommunication networks can help propel a country's economic growth trajectory into a higher sphere, especially in the information age.

The proponents of dismantling the monopolistic regime of the service operator were also helped by swift technological changes. The digitization of telecommunications required huge new investments in the wireline network in order for India to upgrade its telecommunications infrastructure which existing firms were not in a position to make.

The state operator was also unable to invest in cellular technology and offer wireless services to the Indian customers.

In 1991, the government decided that services like cellular telephony, radio paging, e-mail, internet, audio and video-conferencing, V-SAT-based data networks, voicemail, videotex were termed as "Value Added" and made the domain of the private sector (Achar, 1995). In May 1994, the New Telecommunications Policy was adopted, and basic (wireline) telephony was also formally opened up for private sector investment. There are lessons to be learned from both the spectrum auctions and also from the wireline ("basic services") license auctions. We will first examine the latter because the results of using a sealed-bid first-price auction were stark, and counterproductive from the perspective of the country.

### 3.1. Auctions for Basic Wireline Services

In order to auction licenses for wireline services, India was divided into 20 "circles" (or regional divisions) which were then put up for bid to allow one private operator to enter each circle and compete with the government monopoly, the Department of Telecommunications. The applications for licenses to operate in each circle were all supposed to be first-price sealed bids, and the government was explicit how they were to be evaluated. In order to ensure a balanced nationwide growth in telecommunication services especially in the rural areas, the pre-conditions for participating in the license auctions included:

- The private entity had to be a joint company formed with the participation of an Indian company
- Licensees must give at least 10 per cent of all lines to rural areas
- The licensee's network must cover all the districts in the area within 24 months
- Prices charged by the DoT (where it was the competitor) would be ceiling for the prices that private sector firms could charge; of course, they had the freedom to charge a lower rate (Chowdary, 1995)

However, when it came to evaluating the bids at the auctions, the government's motives reflected the need to use the bidding process to generate revenues. Only 15% weight was given to the speed at which the network would be rolled out, 10% was given for rural coverage, and 3% for indigenous equipment used in the network. A "whopping" 72% weight was for the level of the license fee bid (Dokeniya, 1999).

Even so, many foreign telecommunications companies participated in the bidding for the right to offer basic (wireline) telephony in India. The main attraction was the then widely used number of 250 million "middle class" potential customers, and the waiting list of more than 3 million. Companies that bid included multinational like AT&T, US West, Bell Atlantic, Nynex (at that time a separate company), NTT, and Bell Canada, and small ones like Bezeq of Israel, and Shinawatra of Thailand. Their Indian partners included the Tatas, the Birlas, RPG, Reliance, BPL, Essar, Shyam Telecom and Himachal Futurisatic Communications Limited (HFCL).

The first round of the basic service (wireline) auction attracted 80 bids for 40 licenses from 16 companies. In 7 of the circles (Delhi, Haryana, Karnataka, Maharashtra, Orissa, Rajasthan, and Uttar Pradesh West), the bids followed some level of rationality. In circles where HFCL was the highest bidder, their bid was higher than any of the second placed bids, and almost *higher than all of the remaining bids combined* (Dokeniya, 1999). Clearly, because of the problems of insufficient information about an industry that they were planning to enter, their valuations were completely off that of the others, and they were plainly subject to the well known "winner's curse".

The first bidding round unraveled due to a number of issues. Many companies objected to HFCL being the winner in all the circles in which they participated. HFCL also wanted to back out of some of the circles in which they had won for fear of being unable to pay all the license fees. Further, many circles received either one bid or no bids which was considered unsatisfactory. Thus, the government had to call for a second round of bidding for basic (wireline) services.

The second round was also affected by some structural problems. Because the more lucrative (grade A) circles had been awarded in the first round, the only circles let for bid in the second round were the less lucrative circles. In the circles in which HFCL reneged, the government specified a minimum bid which was considered too high by most participants. In addition, caution among the players and the lack of credibility about the outcome of auction lead to only six bids. Thus, at the end of the bidding process, of the 21 circles (largely similar to the state boundaries) that were put up for auctions, only 6 licenses were issued for Andhra Pradesh, Gujarat, Maharashtra, Madhya Pradesh, Punjab and Rajasthan. Currently (circa Auguat 2001), the roll-out of basic wireline telephony has been slow at best.

### 3.2. Spectrum Auctions for Cellular Services

In accordance with the policy espoused by the government in 1991, the Department of Telecommunications initiated the licensing of eight mobile licenses in the four major metropolitan cities of Chennai, Mumbai, New Delhi and Calcutta along with paging services in 26 major cities.

The cellular licenses for the four metropolitan cities were auctioned through a single sealed-bid auction for two licenses in each city (i.e. 8 licenses in total). In such a process, all applicants were allowed to submit one sealed bid to apply for the license. The selection criteria employed for allocating licenses was supposed to be the license, the speed of roll-out, tariffs, reputation and experience of the firms. However, the Indian Metro cellular auction lacked transparency because the selection criteria were not announced publicly, unlike in the *later* case of the basic (wireline) services, before the bids were made (Dokeniya, 1999). Consequently, the winners of the auction seemed subjective at best and arbitrary at worst and the losers took recourse to the courts. It was after a long legal dispute among the participants of the auction and the Department of Telecommunications that the eight licenses were finally awarded in July 1995. As per the terms of the license, the winners had to pay a fixed sum as license fees in the first three years before they switched to paying on a per-line basis.

| Name of Circle | Bidder I | Value of Bid | Bidder II | Value of Bid |
|---|---|---|---|---|
| **Group A** | | ($m) | | ($m) |
| Andhra | JT Mobile | 319 | BPL | 308 |
| Gujarat | Birla AT&T | 571 | BPL | 393 |
| Karnataka | Modi | 444 | (a) Birla | 420 |
| | | | (b) JT Mobile | 420 |
| Maharashtra | Birla AT&T | 528 | BPL | 466 |
| Tamil Nadu | BPL | 266 | Hinduja-HCL | 172 |
| **Group B** | | | | |
| Haryana | Escotel | 78 | Aircell Digilink | 76 |
| Kerala | BPL | 165 | Escotel | 123 |
| Madhya Pradesh | CellularComm | 19 | Reliance | 2 |
| Punjab | Modi | 403 | JT Mobile | 123 |
| Rajasthan | Modi | 122 | Aircell Digilink | 67 |
| UP (East) | Modi | 72 | Koshika | 67 |
| UP (West) | Escotel | 129 | Koshika | 82 |
| West Bengal | Reliance | 13 | No bidder | |
| **Group C** | | | | |
| Arunachal Pradesh | No bids | | | |
| Assam | Reliance | 0.4 | No bidder | |
| Bihar | Koshika | 43 | Reliance | 0.8 |
| Himachal Pradesh | Bharti telnet | 5 | Reliance | 0.4 |
| Orissa | Koshika | 28 | Reliance | 0.8 |
| North-East | Hexacom | 0.5 | Reliance | 0.4 |
| Jammu Kashmir | No bids | | | |

**Table 1: Bidding for Cellular Services in India**

The second round of cellular auctions for 40 licenses in 20 states was held in 1995. Again, there was a single sealed bid auction. And once again, the DoT did not specify the selection criteria involved. However, it was well known that the prime criterion for the award of licenses for cellular, mobile telephony, 2-way paging was the highest license fee (EPW, 2000).

Bidding for the cellular licenses was a gamble because the private companies lacked proper market research on telephone usage in any particular area. The DoT, the only company with data on telecom usage and demand patters made no attempt to share such information with prospective bidders either (EPW, 2000). A further complication was the fact that the Indian partners, by and large, in the telecom venture had no experience in the telecommunications business.

For the government of India, despite the problems of the auctions and the unrealistic high bids, the cellular licenses raised over $7 billion. However, this method of licensing though a single sealed bid auction without any information of the selection criteria has led to many problems in the cellular sector.

There is little doubt that the bidders *overestimated* revenues and demand patterns. However, the high license fees which formed *50 per cent of the total roll-out cost* for a cellular operator led to high tariffs for the consumers. This in turn impacted demand for these services. Moreover, the license fees had to be paid up front every year. That is, it did not matter whether the new operators had a network, subscribers, traffic or revenues, but they had to pay an up front fixed fee to the government every year.

Given the high sunk cost of initial investment, the lower than expected subscriber base and the high license fees, cellular operators in India, with the exception of those in Mumbai and New Delhi markets have been posting losses from the outset. In 1999, the cellular telephone industry was *posting losses of $92 million every month* (Achar, 1999). Industry experts say that overestimation of market size and usage was the main culprit with companies, which had projected 300 minutes of usage, could get subscribers to barely talk for 100 minutes in a month. Besides, average revenue per user was only $23 compared to projections, which had ranged between $41 and $58 per month. As a result by July 1999, the private cellular companies owed almost $900 million in license fees to the government and many had decided to exit the business completely (Achar, 1999).

Given all the problems experienced in both the cellular and wireline liberalization and privatization process, the government of India decided to introduce some changes. In September 1999, inspite of opposition lead by the Congress Party, the BJP government decided to implement a new National Telecom Policy which allowed the private sector operators in the telecommunication service providers to shift from a license fee regime to *a revenue sharing one*. The only caveat, which companies accepted gladly, was that in order to qualify for the new revenue sharing arrangement, they would have to withdraw their pending court cases against the government. All the twenty-nine firms, including 22 cellular operators decided to move to the new arrangement. Of course, the actual level of the revenue being shared has become an *ex poste* negotiated amount based on the company and the region in which it is operating.

After shifting to the revenue sharing regime, which also led to a lower cost for subscriber because of 25 per cent lower rental charges, the cellular market grew significantly between April 1999 and March 2000 by over 80 per cent in terms of revenue and 58 per cent in terms of subscribers. At the end of March 2000, there were 1.88 million cellular subscribers, with an addition of 650,000 during the year. However, there are still 6 slots that are yet to be auctioned off in four different circles.

**4. Some Axioms for Conducting Auctions in Developing Countries**

There are some important lessons to be learned about conducting spectrum auctions in developing countries where the objective and subjective conditions are very different from that in the U.S. Based on many years of experience with development projects, many of these conditions could have been articulated *ex-ante* the conduct of actual spectrum auctions in countries like India. However *ex-poste* the experience, the axioms that we articulate are much more powerful.

First of all, the reasons why most governments in developing countries are auctioning off the airwaves to private telecommunications operators is because these countries do not have sufficient funds to build the wireless networks. Secondly, in most countries, there is very little technology or technical skills to actually design, build and manage telecommunications networks, wireline or wireless. Thus auctions enable the countries to allow global telecommunications companies to enter the country, albeit at a cost for obtaining the "right-of-way". Of course in order to have some control over an infrastructure good, these "foreign" companies are often made to ally with some local entity, or at least obtain a significant amount of materials and human resources from local sources. Thus, after the spectrum licenses are won, there would a need to negotiate the prices and quantities of many inputs going into build the networks with local suppliers.

Thirdly, there is the issue of the governments averting the risk of getting into a reasonably fresh field enterprise in order to safeguard their finances for more immediate necessities like food, water, health, shelter and education. Thus, the telcos, especially if they win will be dealing with a market that is new and unused to telecommunications services. They will have to learn, sometimes while "doing", about the consumers' (residential, business, and commercial) willingness-to-pay and price elasticities for different services. Thus revenue estimates would be very tricky to make. They also have to estimate the cost of the telecom networks with little or no reliable data from the past. Even in countries where there is already a reasonably well established state controlled network, because of subsidies etc., the cost of building the network would not translate easily to a private venture. Thus there would be significant uncertainty about exactly what the value of the spectrum would be, and how this added license cost would impact the profitability of the venture.

It is extremely important for the developing country government to allow fair and transparent auctions. In the situation where the country depends on the technological expertise and financial entrepreneurship of a global telecommunications company, an environment should be created to enable them to participate in the current auction, and any future auctions that may be necessary in order to expand the network. The rules of the auction should be straightforward and clear, including who should participate, what they are bidding for, and when the auction ends.

*Axiom 1: The auction should treat all participants equally and fairly, and should be transparent in both the rules of the auction and the outcome. It is best for the auction rules to be simple.*

It is clear that the governments are using the spectrum auctions to raise money for their exchequers. Thus, an auction that makes sense from the point of view of the developing country government has to have the potential to raise significant license fees. The auction mechanism should enable those participating agents to bid as high as their individual valuation of the spectrum (i.e. reveal true values). This objective is not very different from that in most developed countries around the world as the recent experience in Europe has shown. However, most developing country governments would be happy to trade off incentive compatibility requirements provided that they obtain license fees above what they consider to be true value of the resource.

*Axiom 2: The auction should have the potential to provide as much revenue for the public treasury as the value placed on the resource by the government.*

Even more importantly, and also due to the lack of available funds, most Third World governments would like to have the wireless networks actually implemented in order to expand telecommunications services in their country. Even if one does not take into account the consumption tax that could be collected on communications revenues, it is clear to most governments around the world that their countries need to be part of the telecommunications revolution, and allow their citizens to easily access the Internet. Thus, the auction is considered mostly as a way to provide right-of-way for a serious telco to move forward and build out the network. Indeed, in most developing countries, most bidders have to be pre-qualified to bid in auctions, and most new entrants into the global telecommunications industry are subject to fairly intense scrutiny, frequently to be disqualified.

*Axiom 3: The auction should be just the initial stage in a rapid build-out of the (wireless) telecommunications network.*

Due to the fact that most people in developing countries, by definition, have a low disposable income, the network that is built should be as cost effective as possible. This means that the government should really work in partnership with those who have won the right to build the network in order to keep the cost low. This does not mean that the government should entertain a downward re-negotiation of the outcome of the spectrum auction. In fact, if this were to happen, both axiom 1 and 2 above would get violated. What it means is that the government should enable the winning telco to pursue all activities subsequent to the auction with efficacy and expediency, with a focus on keeping costs low. Indeed, the auction design and winning the license should not lead to an adverse effect on subsequent network cost.

*Axiom 4: The auction design should ensure that the subsequent to its completion that the network could be built and deployed at the lowest possible cost.*

As the experience in India shows, if the auction does not lead to a conclusive outcome expeditiously, then it could be subject to a number of untoward influences. To be clear, in most developing countries, given the amount of money involved in these spectrum auctions, if the conclusions are not known rapidly and transparently, it could be the subject of corruption from even those Ministries involved with designing the auction. Thus, one of the most important considerations in a developing country is to have these auctions conducted in a way that final results can be made public soon.

*Axiom 5: The auction design should ensure a rapid conclusion to the auction.*

**5. Limitations of the Vickrey Second-Price Auction**

Competition: Both too much competition and too little competition affects the outcome of the Vickrey auction from the perspective of the developing country. The most common occurrence in developing countries around the world is too little competition. There are a number of reasons for this, the main one being the general risk aversion of telecommunications companies going to countries where the disposable income is very low. Exacerbating this risk averse behavior are the conditions that are imposed by most countries in order to ensure that local companies are partners in telecommunications ventures. This "local content" rule takes many forms, the most common being the imposition of a certain ownership requirement (usually 51% or more) of any joint venture bidding on telecom licenses. Given this conditionality, there are very few companies in the U.S. and Europe who are willing to be minority partners in telecom ventures, and if they are, there are very few local companies with either the technology capability or execution wherewithal to partner with. The end result is a low level of competition, usually 2-3 companies bidding for the spectrum licenses, and a result that is close to the New Zealand case where the second-price is very low, leading to a bargain for the winning firm and insignificant revenues for the government. In a situation of low competition, the use of a Vickrey auction will violate axioms 2 above, and most likely axiom 1 as well.

Too much competition leads to some counter intuitive behavior. In the sealed-bid Vickrey auction, it is clear that the bidders should bid more aggressively if they are faced with more competition. Every bidder is supposed to have some *private* information about his valuation of the resource (eg., the wireless spectrum), and usually bids something lower than that value, thereby realizing a "profit" (value – bid). As there is more and more competition, to maximize the chances of winning, the bidder needs to trade off the profit, and make bids much closer to the valuation. Thus, too much competition should be good for the auctioneer because of aggressive bidding. However, if the winners end up bidding much more than any one else, then there could be the advent of Winner's curse, leading to the desire to renegotiate the license fee, or slow down the building of the network. Thus, too much competition coupled with over aggressive bidding could lead to conditions where axiom 3 will be violated, and perhaps even axiom 5.

In the case of telecommunications, most bidders have the same information about the marketplace (i.e. who will be their customers, how many customers they will have, and how much they will be willing to pay), and the technology (what does it cost to build out a network, and what is the cost of capital). Thus, one can argue that the resource being auctioned off has *common value* rather than private value. Thus, while different bidders might estimate this value slightly differently, the basis of the estimation has to be common. When faced with significant competition, the person who wins will be the one who *over estimates* the value of the resource. If the bidding is done rationally, the bidders should compensate for this *selection bias* by bidding less aggressively. Thus, rather than being good for the auctioneer, rational behavior in a common value auction which should lead to under bidding, will lower revenue (Rothkopf, 2000). In this case, axiom 2 will be violated.

A further problem with too much competition stems from overhead costs for the auctioneer. With every new bidder, the auctioneer, the developing country government, has to perform due diligence, prepare bid forms and formats, participation mechanisms etc. Thus, the net revenue would be reduced by having to incur the additional administrative costs of too much competition.

Delays and the Potential for Rule Changes: If, in the common value auction, the bidders do not compensate for the selection bias and continues to bid aggressively, the one who wins will realize what we described as "winner's curse". Clearly, the winner would have made the most optimistic estimate of the resource that has a common value to all the bidders. The entity that obtains the license will be extremely concerned about over bidding, as that which happened in India for basic services, leading to either flight or an attempt to renegotiate the bid price for the license. In this case too, the auctioneers revenue will be reduced. Given the time taken for recovering from a winner who "defaults", axiom 5 will certainly be violated. In addition, unless the government is unwavering in it's reluctance to change the license fee, axiom 1 will be violated. If the winner slows down the building of the network because of all these activities, axiom 3 will be violated.

Third Party Negotiations: In the case of spectrum auctions, winning the resource is not the end of the story; the network has to be built. In developing countries, where the bidding is done by international firms (in partnership with local firms), the network would be built by using local third parties who will provide a number of inputs including labor, materials and some technology, the prices of which need to be negotiated. Further, these third parties neither have had a long-term relationship with the winning Telco, nor do they expect to in the future. In every country, there would be a demand to use their own local suppliers for labor and materials. Thus, these third-party negotiations would be a one-shot deal, and the Vickrey auction format could be problematic.

From the arguments outlined in section 4, it is clear that the Vickrey auction needs to be transparent in that all bids, including who made the bids, have to be announced at the end of the auction. The winning party in a Vickrey second-price auction will pay less than their bid, the difference may well be significant. Third parties with whom the Telcos have to negotiate will know the "profit" (bid minus second-price) that the latter obtained from the license auction itself, and will try to extract a considerable amount of it, increasing the final cost of the network (Rothkopf et al, 1990). All of this will lead to the violation of axiom 4.

What should be done, knowing that the third-party negotiations would increase network cost? The auctioneer can use some kind of encryption that limits knowledge about the bids to only the participants of the auction. However, keeping this knowledge secret is extremely difficult in most countries, and would be virtually impossible in a developing country. A number of options are available to bidders, including bid shading (i.e. all bidders simply bid a little less, thus violating incentive compatibility), and inclusion of all third-party suppliers in the bidding consortium (increasing effective cost and lowering net value of the resource). In either case, the auctioneer, i.e. the developing country government, will lose expected revenue.

<u>Bidding with Financial Constraints</u>: Much of the literature on auctions assumes that the bidders care about winning but bid only *up to* their individual valuation of the resource. In the Vickrey second-price auction bidders truthfully bid their actual valuation. In many cases, especially in developing countries, the bidders might not be able to bid their valuation because of financial constraints. Capital markets might allow the bidders to obtain financing for a lower amount than what bidders think the licenses are worth. This financial constraint is usually dynamic in the sense that it may be different at different periods of the evolution of the telecommunications service. At the time of the auction, financial institutions may not be very optimistic about the chances of winning, or more importantly, the chances of successfully (i.e. profitably) implementing a network. As the spectrum licenses are bid, and the network is built out, this risk aversion would get moderated, and further capital outlays would materialize. When the wireless telecom services are actually sold, and as the customer base expands, the financial constraints might get further alleviated.

Thus, a financially constrained Vickrey auction would certainly lead to a number of inadequate outcomes. First, the actual bids would be the financially constraints rather than the value of the resource. In each country, this would reflect overall *country risk* rather than the value of wireless communications. Clearly, the auctioneer, i.e. the developing country government, will lose revenue because of this. Secondly, the winners would be those who are well connected to the capital markets rather than those necessarily able to implement the network effectively. One could argue that the ability to arrange financing is a reflection of the expectations of the capital markets regarding the success of the bidder in executing the project. However, given the recent valuation meltdown of the telecommunications sector, one cannot be sanguine about this. If there is no congruence between those who can arrange financing for the spectrum bids and those who can execute the network, the developing country will lose out in the long run by not having telecommunications service. This will violate axioms 2 and 3, and perhaps even axiom 4.

Value Uncertainty: Models of auctions, including Vickrey second-price auctions assume that the bidders know the value of the resource, for example the spectrum. In the case of auctions in developed countries, one could argue that this assumption is reasonable. The telecommunications market is fairly well developed, and most companies know the price elasticities of the consumers, business and residential, and the size of the markets in each segment. The cost of building out the network is also known with a reasonable degree of certainty. Further, a number of spectrum auctions themselves have been conducted over the years, and the bidders have learned the art and science of valuing wireless spectrum. *As a caveat, however, the 3G auctions in Europe show that the need to obtain "right-of-way" for new technologies will lead to completely unlearned speculative behavior*.

In the case of developing countries, the notion that those participating in spectrum auctions will have a good knowledge of the value of the resource is overly optimistic. If there is uncertainty, then the behavior of the bidder's will not elicit truth revealing that the Vickrey second-price auctions are supposed to. Two of the major problems of conducting a Vickrey auction in a situation with uncertainty are untruthful bidding, and wasteful counter speculation. Risk neutral bidders, i.e. those with linear utility functions, are best off bidding the *expected value* of their valuation of the spectrum in a single-shot private value Vickrey auction. However, those who bid for spectrum in developing countries are mostly risk averse; their utility functions are concave functions of the payoffs. Sandholm (1996) proves that risk averse bidder's with uncertainty about the value of the resource will not bid their expected value in a single-shot Vickrey auction. In fact, in the case of value uncertainty, the optimal bid for a risk averse bidder is to bid less than their estimated expected value. This will lead to two outcomes: a lowering of *expected* revenues for the auctioneer, and the potential of winner's curse because those who win will be the ones who may be least risk averse at the advent of the auction, but who might be unhappy with their bid subsequent to it.

The second result of uncertainty in the value of the resource is "counter speculation" whereby the bid of each agent is based on speculations about what the other bidders would do. It should be recalled that one of the main motivations of the Vickrey auctions was that the dominant strategy for each bidder was to bid truthfully without any consideration for the bids of other participants in the auction. Sandholm (1996) proved that in situations of "local" uncertainty about the value of the resource, even a risk neutral bidder would have the incentive to counter speculate. The main reason for this is that all bidders will try and take some action to reduce the uncertainty in their own (i.e. "local") valuations. Given two different bidders, each of who has a different level of local uncertainty, the optimal amount of information that one of them should collect would depend on how much is needed to get closer to the other. Conversely, rather than collecting this information, the agents bid might compensate for the difference in the levels of information each of them has. Indeed, it is a common practice in most developing country auctions to speculate about whether or not the other bidders have a favored status with the auctioneer, and hence more information about the value of the spectrum or not.

In both cases above, axioms 2 and 3 that we proposed will be violated. In order to address these issues, one needs a dynamic version of the Vickrey auction.

**6. The Vickrey "Share Auction"**

As argued above, the Vickrey second-price auction has a number of limitations for use in spectrum auctions in developing countries including lack of competition, common value of the resource, aggressive bidding, third-party negotiations and value uncertainty. These issues prevail whether one uses a sealed-bid single-shot Vickrey auction, or an ascending price multiple round Vickrey auction. We propose a Vickrey "share auction" that would alleviate most of these problems.

The Vickrey share auction would work as follows: Each bidder would have to bid on the right to use the wireless spectrum, but would bid on the *percentage (i.e. share) of total revenue* that they would provide the government if they won. The winning bid would be the highest share bid, but the winner would only be asked to pay the second highest share bid. Thus, by using the second-highest share, one would get all of the incentive compatibility properties of the Vickrey auction. The question that most bidders would ask is for *how long* will they have to pay a percentage of their revenues? This will bring us to the second part to this share auction: The auctioneer, the developing country government, will have to place *privately* with a trusted authority, say the Chief Justice of the Supreme Court, it's own valuation of how much the spectrum is worth. The winning bidder will continue paying a percentage of its revenue until the government's assessed value of the spectrum is paid off. Thus, time period during which the winning bidder has to pay the license fee would be variable.

Note that this share auction has some of the good properties of the Vickrey auction: There is no incentive for the bidders to bid above their true assessment of what percentage of their revenues they are willing to pay the government each year. Further, making the share bids and the government's valuation of it's property rights public, there will no compromise regarding what each company would expect to make in this telecommunications venture. By having a simple mechanism, the problem of long arduous bidding as in the simultaneous ascending-price multiple round auction is also avoided. In the developing country context, keeping the auction short has significant merits as we have explained in the previous section.

Conditions of axiom 1 will be satisfied because all participants are treated equally and fairly, transparency is maintained, and the rules are kept quite simple. Axiom 2 is also satisfied because the developing country government will obtain the revenue that they expect from the license auction. Clearly, they may not get the maximum amount in a more aggressive competition for the spectrum rights, but they will not feel compromised. The winning bidder only pays when revenue comes in, and the initial license fee as in previous spectrum auctions is not a cost "millstone" even before the network is built. Thus, there will be an incentive to build the wireless network as quickly as possible in order to obtain revenue and soon, get the government "off it's back". After the government's valuation is paid off, the total revenue is completely for the winning bidder. Thus, axiom 3 is also satisfied. Axiom 4 is *easily* satisfied. Third party suppliers do not know how much money is "left on the table" as in Vickrey auctions that are based on price alone. It will be difficult for the local suppliers to increase the price of inputs going into the network, thus keeping the network implementation cost low.

One could say that this share auction is equivalent to having a price auction, and then the winning bidder and the government deciding on a payment schedule over time. Indeed, this is what is happening currently in India's spectrum and wireline auctions. However, allowing the winning bidder and the government to negotiate a payment schedule after the auction has ended may allow corrupt practices to prevail, violating axiom 5. Further, the losing participants could claim that if they knew that the payment schedule could be negotiated, that they would have bid differently. If the perception was one of unfairness, then axiom 1 would certainly be violated.

How does a bidder decide what to bid as the percentage of their revenues they will have to provide the government? The bidders will have to do the exact calculations that they performed in order to bid for the price auction. The will have to forecast the size of the market in terms of revenues that they can generate each year for the project life (usually 30 years). They will have to estimate the cost of building the network without the license fee. They will also have to make a judgment about reasonable rates of return on this project over the project life. Given these numbers, in the price auction they would back-out the maximum license fee that they would be willing to pay in order to ensure the rate of return. In the share auction, the calculations are the same except that instead of an upfront fixed cost of the licenses, they will have to include a percentage reduction of revenue each year for the project life. Any MBA with a spreadsheet should be able to see the equivalence of these two calculations, and back-out the maximum share that the bidder would be willing to pay.

## 7. Conclusions

Developing countries engaged in the deregulation of their telecommunications industry would be tempted to use some form of the simultaneous ascending multiple-round (SAMR) auction to assign spectrum licenses. SAMR auction seem to be fair, generate revenue for the government, reveal firm's estimates of license values, and assign licenses quickly compared to other methods like hearings, lotteries, and sealed-bid auctions, whether first-price or Vickrey second price auctions. SAMR auctions can also be designed to incorporate a wide range of public-policy goals as observed in New Zealand and the U.S. spectrum auctions.

However, this will require firms to correctly compute the value of the differing licenses for their situation. In countries where it is difficult for firms to assess the value of having licenses due to poor market history (lack of data) may make allocation by auction risky since firms that have incorrectly valued licenses may receive licenses that they cannot use efficiently. The assumption is that firm bids do represent values and therefore ability to use spectrum efficiently. Blindly using auctions can lead to very bad outcomes as witnessed in the early airwaves license auctions in Australia and New Zealand and so painstaking attention must be paid to all aspects of auction design and use, but there are some great precedents to make this process a lot easier.

In this paper, we have suggested the use of a Vickrey "share auction" which takes into account the important axioms of auctioning spectrum in developing countries such as equal treatment of the bidders and fairness, revenue generation for the government, rapid build-out of the network, managing network cost after the auction is completed, and a rapid conclusion to the auction itself. Further work is necessary to provide a mathematically rigorous way of proving that, in theory, the Vickrey share auction would be superior to the Vickrey second-price auction.

In a related paper that looked at natural resource concessions, such as oil and mineral concessions, Anandalingam (1987) showed, using mathematical bargaining theory, that a *share* contract would lead to the government (principal) obtaining a greater share of the property value under conditions of greater competition, and a willingness to share in production costs. Indeed, the share auction for spectrum allocation would make the government a partner interested in the success of the telecommunications venture. Given that the most important thing for a developing country is to have communications services and to become plugged into the Internet, anything that results in success should be intensely encouraged.

Vickrey Auctions for Spectrum in Developing Countries                                Anandalingam
**References**

Achar, A. (1999), "Untangling a Telecommunications Revolution", *Telecommunications Journal*, October.

Achar, A (1995), "Reforming Telecoms the Indian Way", *Telecommunications Journal*, October.

Agorics Inc. (1996), "Going, going, gone! Survey of auction types", http://www.agorics.com/agorics/auctions/bibliography.html

Anandalingam, G. (1987), "Asymmetric Players and Bargaining for Profit Shares in Natural Resource Development", *Management Science*, vol. 33, no. 8, pp 1048-1057.

Anandalingam, G., P. Bagchi, and R. Kwon (2000), "Models for Efficient Spectrum Allocation in India", Working Paper, University of Pennsylvania, October.

Chowdary, T. H. (1995), "India: Reforming Telecoms the Indian Way", T*ele-communications* (International Edition), October, pp111-114.

Cramton, Peter C., The FCC Spectrum Auctions: An Early Assessment, *Journal of Economics and Management Strategy*, Vol. 6, No.3, pp.431-495, 1997

Cramton, Peter C., The Efficiency of FCC Spectrum Auctions, *Journal of Law and Economics*, Vol.41, pp.727-736, Oct. 1998

Dokeniya (1999), "Reforming the State: Telecommunications Liberalization in India", *Telecommunications Policy*, vol. 23, pp 105-128.

EPW (2000), "Telecommunications Demonopolisation: Policy or Farce", *Economic and Political Weekly*, February 11.

FCC (2001) http://www.fcc.gov/wtb/auctions/

Fritts, B. (1999), "Private Property, Economic Efficiency and Spectrum Policy in the Wake of the C Block Auction",

McMillan, John, Selling Spectrum Rights, Graduate School of International Relations and Pacific Studies, University of California, San Diego, La Jolla, Ca USA, 1995

Myerson, R. (1981), "Optimal Auction Design*", Mathematics of Operations Research,* vol. 6, pp 58-73.

New Zealand (2001), http://auction.med.govt.nz/
23